\documentclass{article}

\usepackage{arxiv}

\usepackage[utf8]{inputenc} 
\usepackage[T1]{fontenc}    
\usepackage{hyperref}       
\usepackage{url}            
\usepackage{booktabs}       
\usepackage{amsfonts}       
\usepackage{nicefrac}       
\usepackage{microtype}      
\usepackage{graphicx}
\usepackage{amsmath}
\usepackage{amssymb}
\usepackage{lineno}
\usepackage{hyperref}
\usepackage{listings}
\usepackage{subfigure}
\usepackage{comment}
\usepackage{enumitem}
\usepackage[dvipsnames]{xcolor}

\title{On the implementation of large-scale integral operators with modern HPC solutions - Application to 3D Marchenko imaging by least-squares inversion}

\author{
  Matteo Ravasi \\
  Equinor ASA\\
  Bergen, Norway \\
  \texttt{matteoravasi@gmail.com} \\
   \And
 Ivan Vasconcelos \\
  Department of Earth Sciences\\
  Utrecht University\\
  Utrecht, The Netherlands \\
  \texttt{i.vasconcelos@uu.nl}}
  
\begin{document}

\chead{Large-scale integral operators with modern HPC solutions}

\maketitle

\begin{abstract}
  Numerical integral operators of convolution type form the basis of most wave-equation-based methods for processing and imaging of seismic data. As several of these methods require the solution of an inverse problem, multiple forward and adjoint passes of the modelling operator must be performed to converge to a satisfactory solution. This work highlights the challenges that arise when implementing such operators on 3D seismic datasets and it provides insights into their usage for solving large systems of integral equations. A Python framework is presented that leverages libraries for distributed storage and computing, and provides an high-level symbolic representation of linear operators. To validate its effectiveness, the forward and adjoint implementations of a multi-dimensional convolution operator are evaluated with respect to increasing size of the kernel and number of computational resources. Our computational framework is further shown to be suitable for both classic on-premise High-Performance Computing and cloud computing architectures. An example of target-oriented imaging of a 3D synthetic dataset which comprises of two subsequent steps of seismic redatuming is finally presented. In both cases, the redatumed fields are estimated by means of least-squares inversion using the full dataset as well as spatially decimated versions of the dataset as a way to investigate the robustness of both inverse problems to spatial aliasing in the input dataset. We observe that less strict sampling requirements apply in three dimensions for these algorithms compared to their two dimensions counterparts. Whilst aliasing introduces noise in the redatumed fields, they are however deprived of the well-known spurious artefacts arising from incorrect handling of the overburden propagation in cheaper, adjoint-based redatuming techniques.
\end{abstract}

\section{Introduction}
Multi-dimensional integral operators are ubiquitous in almost every form of wave-equation based processing of multi-channel seismic data; they appear in demultiple algorithms, interferometric calculations, and are a key component of many imaging techniques. Their numerical implementation have been investigated in the past, mostly with the aim of applying Surface-Related Multiple Elimination (SRME) to 3D datasets \cite{dragoset2010, pica2005}. Recent research trends have highlighted how SRME can be seen as the adaptive summation of the first two terms of a Neumann series, which is function of the sought reflection response deprived of surface-related multiples and the full-wavefield dataset itself (see \cite{vangroenestijn2009} or \cite{lopez2015a} for a detailed derivation). A more accurate way to solve the demultiple problem requires the solution of the underlying inverse problem by means of a gradient-based iterative solver; in practice, the main advantage of the latter approach with respect to vanilla SRME is that the step of adaptive subtraction is no longer required and gaps in the acquisition geometries can also be handled \cite{lopez2015a}. On the other hand, the computational cost of the overall process is dramatically increased as the most computationally expensive operation -- i.e., Multi-Dimensional Convolution (MDC) between the input dataset and the current estimate of the primary-only data -- has to be performed multiple times. 

In the last decade, a variety of other algorithms that experience similar computational challenges have been developed. All of them are united by the fact that their modelling operator contains one or more MDC operations and an inverse problem must be solved. Examples of such algorithms, categorized on the task they aim to solve, are:
\begin{itemize}
  \item Inversion-based free-surface demultiple: Estimation of Primaries by Sparse Inversion (EPSI - \cite{vangroenestijn2009, lin2013}) and Closed-loop SRME \cite{lopez2015a, lopez2015b};
  \item Deconvolution-based free-surface demultiple \cite{amundsen2001a, ravasi2015}
  \item Marchenko-based processing and imaging -- among others, redatuming via Marchenko equations \cite{vanderneut2015, dukalski2018}, joint redatuming and wavelet estimation via Marchenko equations \cite{becker2018}, Rayleigh-Marchenko equations \cite{ravasi2017}, scattering-Marchenko equations \cite{vasconcelos2019}, internal multiple attenuation in data domain \cite{zhang2019}, and time-lapse redatuming via Marchenko equations \cite{haindl2018};
  \item Interferometric redatuming: interferometric redatuming by multi-dimensional deconvolution \cite{wapenaar2010, wapenaar2011, vanderNeut2013} and target-oriented redatuming \cite{vanderneut2017, vasconcelos2017};
  \item Full-wavefield modeling and inversion \cite{berkhout2011, davydenko2016} and Joint-migration-inversion \cite{staal2013}.
\end{itemize}

As an iterative scheme is required to solve the underlying inverse problem, the adjoint of the modelling operator must also be implemented and special care has to be given to its structure and implementation efficiency. This calls for possible modifications of some of the ideas originally developed for the efficient implementation of 3D SRME to be applicable to the adjoint operators. For example, \cite{lopez2015b} have shown that to be able use the 3D general surface multiple prediction (GSMP) method of \cite{bisley2005} in closed-loop SRME, an extension of the basic concept is required to include also correlation (which is required in the adjoint step). Moreover, whilst SRME relies on a final step of adaptive summation that can compensate for inaccuracies arising during the prediction step, even more care must be taken when implementing these inversion-based methods to avoid that inaccuracies in the forward and adjoint passes may lead to instabilities in the inverse process. Trading computational efficiency for accuracy may be required in these scenarios, and greedy evaluations of the multi-dimensional integrals of convolution (and correlation) type should be applied with more care.

In this paper, we analyse the computational challenges that arise when implementing large multi-dimensional convolutional operators on 3D seismic datasets and present a framework that allows for their evaluation over distributed computer systems. We further discuss how various open-source libraries in the Python ecosystem can be combined to create an efficient end-to-end computational framework that is easy to modify and use for the implementation of a variety of different algorithms. The effectiveness of our implementation is evaluated by applying the forward and adjoint MDC to kernels of increasing size. We show that both operations scale with respect to computational resources and can be deployed over a large variety of HPC architectures. Finally, the proposed computational framework is used to perform full-wavefield target-oriented imaging of a 3D synthetic dataset which contains up to 9801 sources and receivers. First, the so-called Marchenko equations \cite{Wapenaar2014a, vanderneut2015} are solved to redatum receivers from the acquisition level to at a certain subsurface datum. Second, we perform source-side redatuming by means of multi-dimensional deconvolution of the up- and down-going Green's functions estimated by the previous algorithm. Both the full dataset as well as three spatially decimated versions of the dataset are used to investigate the robustness of these redatuming methods to spatial aliasing in the input dataset. Complementing the work of \cite{jia2018, lomas2019, brackenhoff2020}, our study shows that the Marchenko equations can also be solved via direct inversion in three dimensions (and not only via Neumann iterations). Moreover, our example shows how the estimated Green's functions can be further used for target-oriented imaging purposes also in cases where the sampling requirements are mildly violated. Similar to previously published studies performed in two dimensions, we show that downsampling of the input dataset along one of the spatial directions does introduce coherent noise in the estimated Green's functions and reflectivities; nevertheless, the accuracy of the redatuming processes when it comes to the handling of the complex propagation in the overburden is not fully compromised even in cases where the source and receiver spacing are one and a half larger than a quarter of the dominant wavelength of the input focusing functions.

\section*{Theory}
In this work, we are concerned with integral operators of the following kind:
\begin{equation}
\label{eq:integraltime}
g(t, \mathbf{x}_B, \mathbf{x}_A) = \int_{\mathbb{R}} K(t - \tau, \mathbf{x}_B, \mathbf{x}_R) f(\tau, \mathbf{x}_R, \mathbf{x}_A) d\mathbf{x}_R \quad 
\end{equation}
where $K(t, \mathbf{x}_B, \mathbf{x}_R)$, $f(t, \mathbf{x}_R, \mathbf{x}_A)$ and $g(t, \mathbf{x}_B, \mathbf{x}_A)$ are the so-called \textit{integral kernel operator}, the input, and output functions in time-space domain, respectively. Here $t$ is used to indicate time axis, and $\mathbf{x}_B$, $\mathbf{x}_A$, and $\mathbf{x}_R$ refer to the spatial coordinates of a point in three-dimensions, with the latter ($\mathbf{x}_R$) corresponding to the integration support $\mathbb{R}$ of the spatial integral.

As generally neither $K$ nor $f$ have a compact support in time domain (i.e., filters with short time response), equation \ref{eq:integraltime} can be equivalently, and more efficiently, implemented by using its frequency domain representation:
\begin{equation}
\label{eq:integral}
g(t, \mathbf{x}_B, \mathbf{x}_A) = \mathcal{F}_{\omega_{max}}^{-1} \left( \int_{\mathbb{R}} K(\omega, \mathbf{x}_B, \mathbf{x}_R) \mathcal{F}_{\omega_{max}} \left( f(t, \mathbf{x}_R, \mathbf{x}_A \right)) d\mathbf{x}_R \right) \quad 
\end{equation}
where $K(\omega, \mathbf{x}_B, \mathbf{x}_R)$ is the integral kernel operator in the frequency-space domain. In our notation $\mathcal{F}$ and $\mathcal{F}^{-1}$ represent the forward and inverse Fourier transforms, $\omega$ is the angular frequency and $\omega_{max}$ is used to indicate that the output of the forward Fourier transform is truncated to contain only frequencies where the signal spectrum resides. At this point we note the requirement to move back and forth between time and frequency domains arises under two circumstances: i) the modelling operator performs any operation to the output function in time domain (e.g., windowing) as in the case of the Marchenko equations discussed in the Application section, ii) a regularization (or preconditioning) term, used to stabilize the inversion, is required to act on the input function in time domain (e.g., enforcement of time causality as used by \cite{vanderneut2017} in target-oriented imaging applications). In the case where neither of the above is verified, equation \ref{eq:integral} can be further simplified and evaluated directly in the frequency domain.

Numerical implementation of equation \ref{eq:integral} requires a discretisation of the integration domain: to this end, $K$ becomes a 3-dimensional tensor of size $(n_{\omega_{max}} \times n_{\mathbf{x}_{B}} \times n_{\mathbf{x}_{R}})$, $f$ and $g$ are tensors of size $(n_t \times n_{\mathbf{x}_{R}} \times n_{\mathbf{x}_{A}})$ and $(n_t \times n_{\mathbf{x}_{B}} \times n_{\mathbf{x}_{A}})$, respectively. The integral reduces to a matrix-matrix multiplication (or matrix-vector multiplication in the special case of $n_{\mathbf{x}_{A}}=1$) repeated for each frequency slice, also referred to as \textit{batch matrix multiplication} in this work. Moreover, the integration step ($d\mathbf{x}_R$) has to be considered: this is especially important for the case of an irregularly sampled domain $\mathbb{R}$, where Voronoi tessellation can be used to identify the areal extent of each point and a variable scaling is applied to the columns of each frequency slice of the kernel. Such constant (or receiver-dependant) scaling is simply pre-multiplied to the kernel prior to applying batch multiplication as shown in Figure~\ref{fig:implem}.

The forward and adjoint operations of equation \ref{eq:integral} can be written in a compact notation as: 
\begin{equation}
\label{eq:foradj}
\mathbf{g} = \mathbf{F}^H \mathbf{I}_{\omega_{max}}^H \hat{\mathbf{K}} \mathbf{I}_{\omega_{max}} \mathbf{F} \mathbf{f} = \textbf{C} \mathbf{f} \qquad \mathbf{f} =  \mathbf{F}^H \mathbf{I}_{\omega_{max}}^H \hat{\mathbf{K}}^H \mathbf{I}_{\omega_{max}} \mathbf{F} \mathbf{g} = \textbf{C}^H \mathbf{g}
\end{equation}
where $\mathbf{F}$ and $\mathbf{F}^H$ represent the operators performing forward and inverse Fast Fourier Transforms along the time/frequency axis, $\hat{\mathbf{K}}$ is the operator performing batch matrix multiplication with the already scaled kernel ($K(\omega, \mathbf{x}_B, \mathbf{x}_R) * d\mathbf{x}_R$), and $\mathbf{f}$ and $\mathbf{g}$ contain the input and output functions unwrapped into vectors of size $(n_{\omega_{max}} n_{\mathbf{x}_R}  n_{\mathbf{x}_{A}} \times 1)$ and $(n_{\omega_{max}} n_{\mathbf{x}_{B}} n_{\mathbf{x}_{A}} \times 1)$, respectively. Finally, $\textbf{C}$ and $\textbf{C}^H$ are used to compactly define the overall forward and adjoint operators. Given the symmetry of the Fourier operators, the overall forward and adjoint operators are both implemented by first applying the real forward Fourier transform ($\mathbf{F}$) to the input vector, truncating the frequencies up to $n_{\omega_{max}}$ ($\mathbf{I}_{\omega_{max}}$), performing a batch matrix multiplication with either the kernel or its transpose and complex conjugate ($\mathbf{K}$ or $\mathbf{K}^H$), padding the output to the number of frequencies required by the Fourier transform ($\mathbf{I}_{\omega_{max}}^H$ --- noting that the adjoint of a truncation operator is a zero-padding operator) and finally applying the inverse real Fourier transform ($\mathbf{F}^H$). 

From a mathematical point of view, the batch matrix multiplication operator can be equivalently seen as a block-diagonal matrix with each frequency slice of the kernel $K$ representing a block along it main diagonal, i.e. $\mathbf{K} = diag\{ \mathbf{K}(\omega_1), \mathbf{K}(\omega_2), ..., \mathbf{K}(\omega_{N_\omega})\}$ where each frequency slice is a dense matrix of size $N_B \times N_R$ laid out as follows:

\begin{equation}
\label{eq:kernelslice}
\textbf{K}(\omega) = 
\begin{bmatrix}
   K(\omega, \mathbf{x}_{B_1}, \mathbf{x}_{R_1}) &
   K(\omega, \mathbf{x}_{B_1}, \mathbf{x}_{R_2}) & ... &
   K(\omega, \mathbf{x}_{B_1}, \mathbf{x}_{R_{N_R}}) \\
   ... & ... & ... & ... \\
   K(\omega, \mathbf{x}_{B_{N_B}}, \mathbf{x}_{R_1}) &
   K(\omega, \mathbf{x}_{B_{N_B}}, \mathbf{x}_{R_2}) & ... &
   K(\omega, \mathbf{x}_{B_{N_B}}, \mathbf{x}_{R_{N_R}})
\end{bmatrix}
\end{equation}

In this paper we refer to this operation as batch matrix multiplication as it more clearly highlights the independency of the different frequency components of the kernel $K$ in the application of both the forward and adjoint operations. 

\section{Implementating the batch matrix multiplication operator}
Batch matrix multiplication represents the most expensive operator in the chain of operations in equation \ref{eq:foradj}, as it requires acting on the entire dataset used as the kernel. When devising an implementation strategy for such an operator, we must consider the following constraints:

\begin{enumerate}
  \item The kernel $K$ may not fit into a single computer's main memory at one time. As such, an \textit{out-of-core} implementation is required, and the kernel is preferably distributed across multiple compute nodes to minimize I/O operations;
  \item Repeated evaluations of the forward and adjoint operations are required to solve an inverse problem, and;
  \item $n_{\mathbf{x}_B} \ge n_{\mathbf{x}_R}>>n_{\mathbf{x}_A}$ and $n_{\mathbf{x}_A} \ge 1$ - i.e., $K$ is a 3-dimensional array while $f$ and $g$ can be either 2- or 3-dimensional, but they are both generally smaller than the kernel. 
\end{enumerate}

\begin{figure}
  \includegraphics[width=\textwidth]{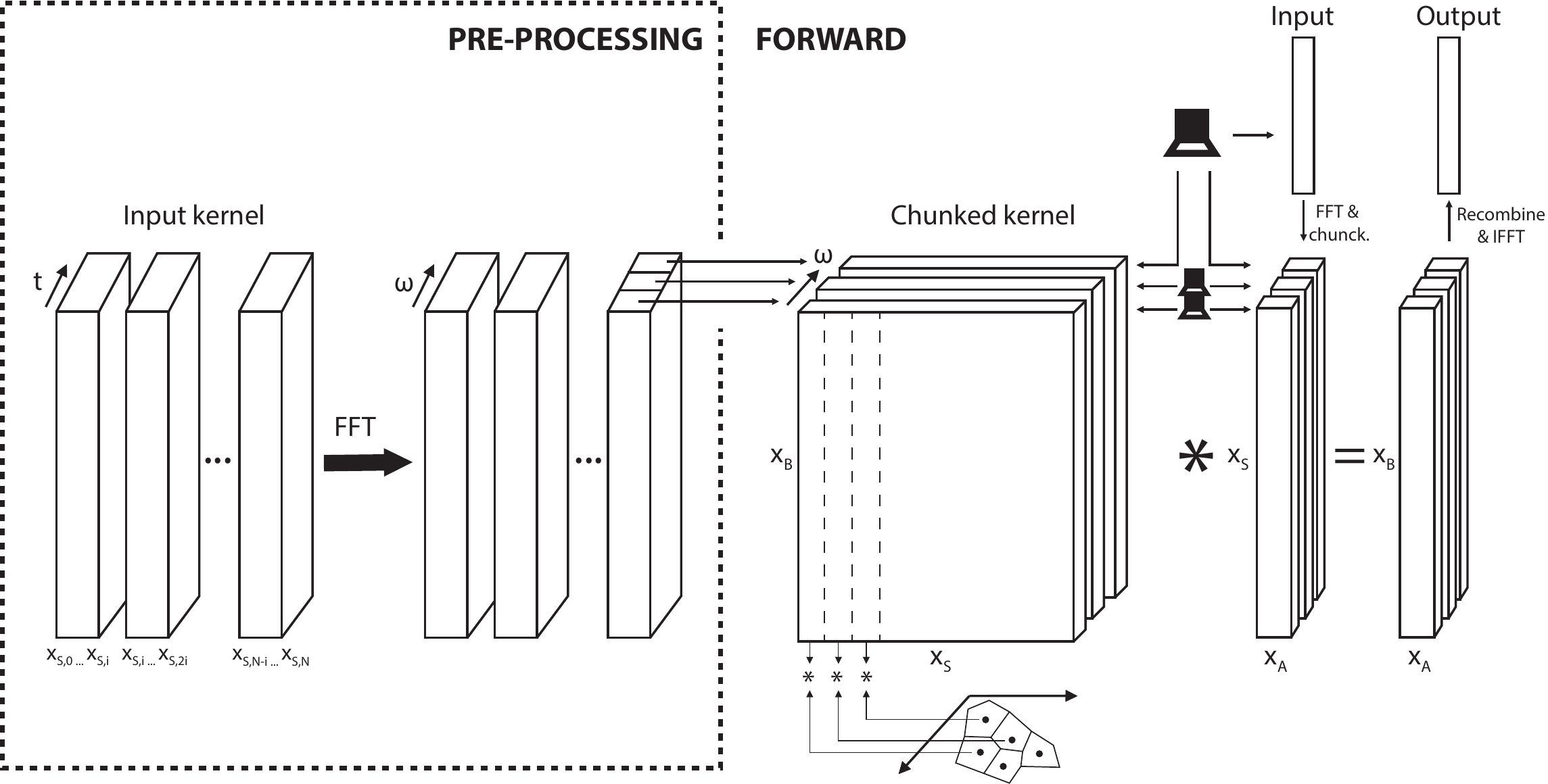}
  \caption{Schematic representation of the pre-processing of the kernel operator and forward pass.}
  \label{fig:implem}
\end{figure}

Figure \ref{fig:implem} presents a schematic representation of the pre-processing and forward pass of the operator in equation \ref{eq:foradj}. In seismic applications, the kernel $K$ is generally available in the time-space domain and originally stored in multiple files containing one or more shot (or receiver) gathers. As solving an inverse problem requires access to the input dataset multiple times, a pre-processing step is usually required to transform the kernel into frequency-space domain. Two alternative strategies can be followed: the input data is read by a distributed system of compute nodes and Fourier transformed to the frequency domain. The different nodes exchange with each other different frequency batches such that ultimately every node has access to a set of frequencies for all sources and receivers in the input dataset. Alternatively, in a pre-processing step, the same group of compute nodes reads a set of source (or receiver) gathers in time domain, converts them to frequency domain and saves them back into a new set of files partitioned along the frequency axis. Whilst the latter approach does obviously lead to an overhead in terms of storage, it is the preferred solution as it guarantees that the time-to-frequency transformation of the entire input dataset is performed only once upfront. The former approach, although more appealing from a storage point of view, comes with the risk of losing the frequency domain representation of the input data in case of hardware failure. This also requires performing the same transformation multiple times if several inverse problems are solved independently from each other as further discussed in the Discussion section.

\begin{figure}
  \includegraphics[width=\textwidth]{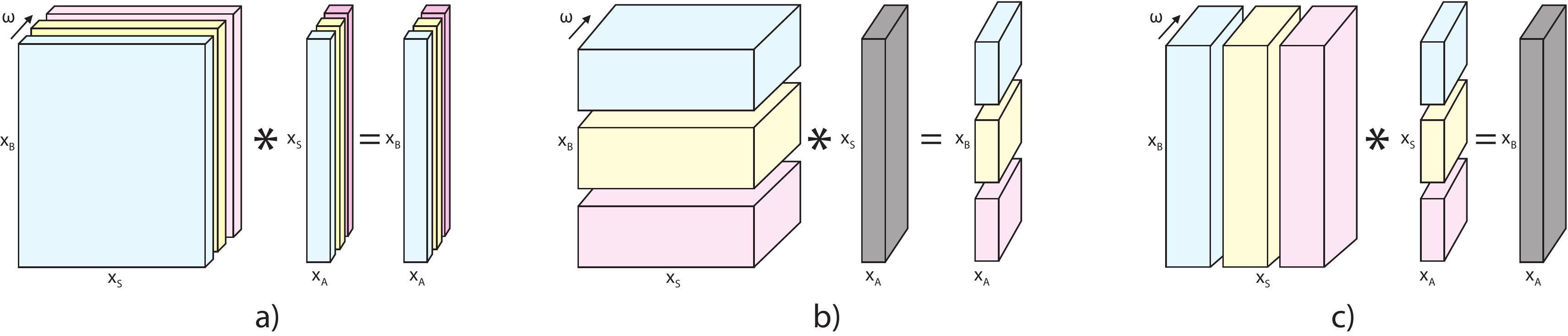}
  \caption{Schematic representations of the possible memory layouts of the 3-dimensional kernel operator, input and output vectors where chunking is performed over i) frequency slices, b) rows, and c) columns. Blue, yellow and red correspond to different compute nodes where part of the kernel resides, while grey is used to indicate that the input (or output) vector needs to be available to all compute nodes.}
  \label{fig:matrixmult}
\end{figure}

The choice of partitioning the input dataset over the frequency axis is justified by the fact that we wish to to compute both the forward and adjoint passes of the MDC operator (\textit{constraint 2}). As the core computation for each frequency slice is a matrix-matrix multiplication, chunking the kernel along its row space (i.e., integration axis) would only be favourable during the forward pass where each chunk is responsible for computing a group of values in the output vector (Figure \ref{fig:matrixmult}b). This is however not the case for the adjoint pass, because each chunk can only be used to partially compute the elements of the output vector and communication across compute nodes is required to sum their partial outputs. The opposite scenario occurs when chunking is performed along the column space (Figure \ref{fig:matrixmult}c). On the other hand, we can take advantage of the independency of different frequencies to avoid any data transfer during the batch matrix multiplication step (Figure \ref{fig:matrixmult}a). By performing the chunking along the frequency direction, no data transfer is required for those applications performed fully in the frequency domain and only at the end of each batch matrix multiplication for those applications where the output vector needs to be converted back to time domain. This way, the kernel is distributed only once at the beginning of the computations and different chunks never leave the compute node to which they have been initially assigned.

Finally, in geophysical applications the kernel in time-space domain is a generally a real valued function. The orthogonal Fast Fourier transform (FFT) for real inputs should be used to implement the Fourier operator $\mathbf{F}$. In this context orthogonal simply means that both forward and inverse transforms are scaled by $1/\sqrt{n_t}$. Doing so, the adjoint of the FFT operator is equivalent to the inverse FFT. Moreover, using the FFT for real signals, only positive frequencies must be computed and stored.

\section*{Open-source software stack}
The computational framework proposed in this paper aims at creating an efficient and scalable implementation of numerical integrals of convolution (and correlation) type. At the same time, it provides users with an easy-to-use interface that resembles as much as possible the underlying mathematical equations to be solved. To achieve this, the following three key ingredients are required: i) an efficient storage format for large N-dimensional regular arrays, ii) the possibility to distribute computations over a cluster of compute nodes, and iii) the ability to define modelling operators and solve inverse problems seamlessly from small scale two-dimensional examples running on a single general-purpose computer to large three-dimensional field scale applications.

In this work we take advantage of the expressivity of the Python programming language and the availability of a large variety of open-source libraries in the Python ecosystem. The \textit{Zarr}\footnote{https://zarr.readthedocs.io/} file format is chosen to store the kernel of the multi-dimensional convolution operator; this storage format is selected because it allows chunked, compressed, N-dimensional arrays to be written and read concurrently from multiple threads or processes. Concurrent reading and writing is achieved by storing the entire N-dimensional array in multiple files, containing N-dimensional chunks of the entire data, alongside with a metadata file enabling correct interpretation of the stored data. Moreover, Zarr provides classes and functions for working with N-dimensional arrays that behave like NumPy arrays but whose data is divided into chunks and each chunk is compressed. 

The \textit{Dask}\footnote{http://dask.org/} library is used to enable distributed computations. Parallelism in Dask is achieved by means of a task scheduler that builds dependency graphs and schedules tasks to a pool of workers. Whilst Dask allows parallelism at different levels -- from threads/processes on a single compute machine to distributed computing over a group of machines -- our application targets the latter lever of parallelism as it ultimately allows the kernels of our integral operators to reside in memory at any time during computations, split over different compute nodes. Being able to limit I/O operations to a minimum provides obvious benefits in any HPC system, and this is of particular importance for cloud computing environments -- more details are provided in the Discussion section. Additionally, Dask provides distributed data structures with APIs similar to NumPy arrays (or Pandas Dataframes) and eases the loading and distribution of large input dataset in a variety of formats including Zaar. 

Lastly, the \textit{PyLops} framework (Ravasi and Vasconcelos, 2019) is used to provide a high-level symbolic representation of linear operators, easing the setup and solution of inverse problems. Whilst initially developed for in-memory computations, the ability of Dask to transparently handle distributed N-dimensional arrays as if they were in-memory Numpy arrays allowed us to easily adapt the PyLops framework to out-of-core operations where Dask arrays and its API are used as a drop-in replacement for Numpy routines and arrays. In this work operators and solvers from the \texttt{pylops-distributed}\footnote{https://pylops-distributed.readthedocs.io/en/latest/} library are therefore used to solve large systems of equations like those discussed in the Application section.

\section*{Benchmarking the Multi-Dimensional Convolution operator}
\begin{figure}
  \centering
  \subfigure[]{\includegraphics[width=\textwidth]{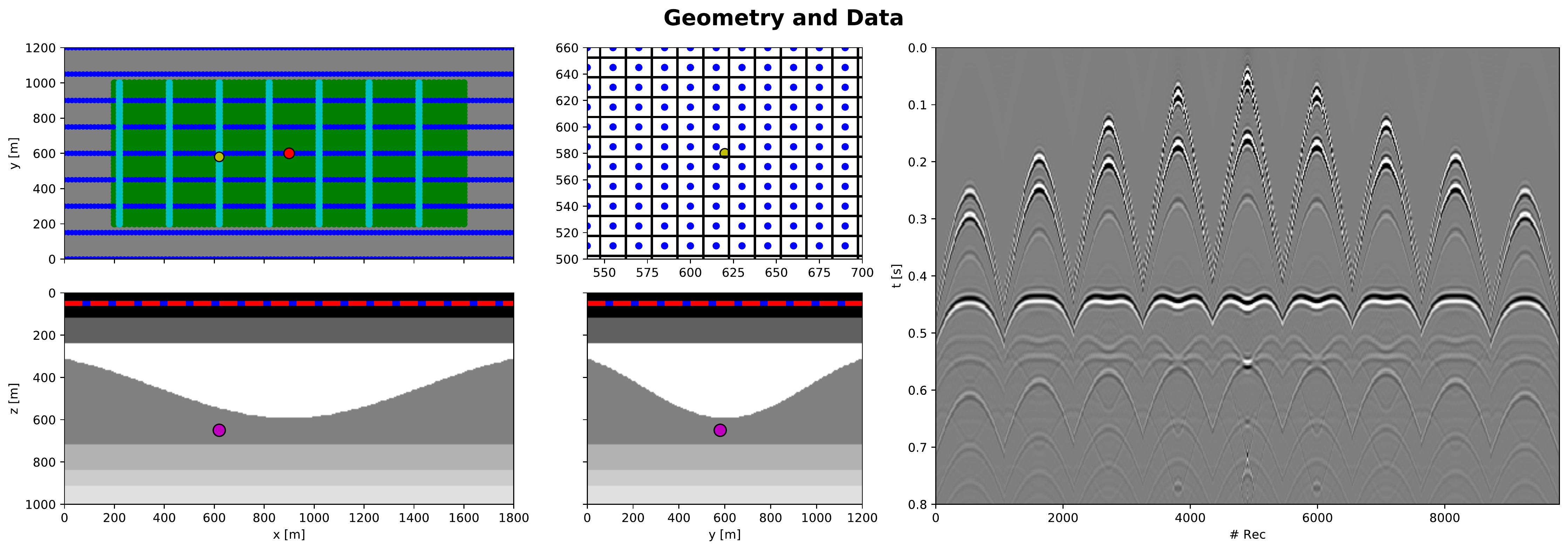}\label{fig:geom}}
  \subfigure[]{\includegraphics[width=\textwidth]{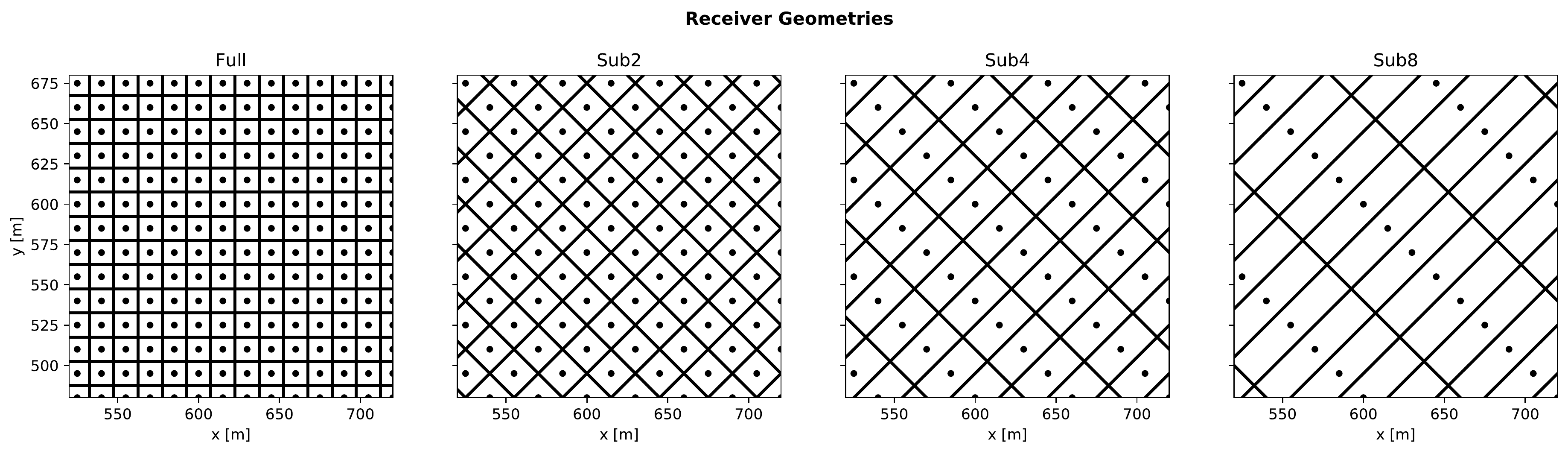}\label{fig:geom1}}
  \caption{Model and data for numerical example. a-b-c) Slices of density model along x, y, and z axes, respectively. Sources are represented by red dots in top view (and lines in section views), receivers are indicated by blue lines and a purple dot is used to indicated the subsurface point. d) Close-up of acquisition geometry around subsurface point and Voronoi tessellation used to define the areal extent $d\textbf{x}_S=900 m^2$ for the MDC integrals (subsampling=4 used for this figure). e) Reflection data for source in the middle of the model and 3 lines of receivers.}
  \label{fig:geom,geom1}
\end{figure}

We consider the application of forward and adjoint operations in equation \ref{eq:foradj}, using a seismic reflection response as the kernel of the integral operator. The constant velocity ($c=2400 m/s$), variable density model in Figure \ref{fig:geom} is used to generate the input dataset. The acquisition geometry consists of a regular grid of 9801 co-located sources and receivers. The dataset is generated using a staggered-grid finite difference modelling scheme, transformed to the frequency domain, truncated to contain the first $n_{\omega_{max}} = 300$ frequencies ($f \leq 73 Hz$), and stored in the Zarr file format. Additionally, we create 3 subsampled versions of the original dataset, whose source (and receiver) geometries are shown in \ref{fig:geom1}.

Following the theoretical requirements for the kernel to be used in the Applications section, each frequency slice of the kernel is a $N_S \times N_R$ matrix, whose \textit{ij}-th element is equal to $\hat{v_z}(\omega, \mathbf{x}_{S_i}, \mathbf{x}_{R,j}) = v_z(\omega, \mathbf{x}_{S_i}, \mathbf{x}_{R,j}) / S(\omega)$. Each element represents the particle velocity recording at the j-th receiver from the i-th monopole source ($v_z$), deconvolved by the source signature $S(\omega)$. The input vector $\textbf{f}$ is instead created by numerical modelling of a single event from a subsurface point $\textbf{x}_A=(580,620, 650) m$, convolved with a Ricker wavelet ($f_{dom}=20Hz$). Forward and adjoint modelling are performed for the following combinations of data and compute resources (Table \ref{table:1}). 

The first benchmark is performed on compute nodes equipped with 2 octa-core Intel(R) Xeon(R) CPUs @ 2.90GHz, and 128GB of RAM each. Compute nodes are connected to each other via the \texttt{SSHCluster} functionality of Dask. Moreover, the Anaconda Python distribution is used to take advantage of distributions of the NumPy and SciPy libraries with the Intel Math Kernel Library. Two additional benchmarks are performed using different cluster configurations, namely:
\begin{itemize}
  \item An HPC cluster with Portable Batch System (PBS) job scheduling using the \texttt{PBSCluster} Dask interface;
  \item A Kubernetes cluster on Microsoft Azure Cloud with \textit{Standard\_E16\_v3} machines. The \texttt{dask-kubernetes} library is used to deploy the Dask scheduler and workers in the cloud environment;
\end{itemize}
In both cases, the number of CPU and RAM visible to Dask is limited to 12 and 128GB, respectively. This ensures that the compute nodes have as similar as possible specifications to those used in the first experiment. 

\begin{figure}
  \includegraphics[width=\textwidth]{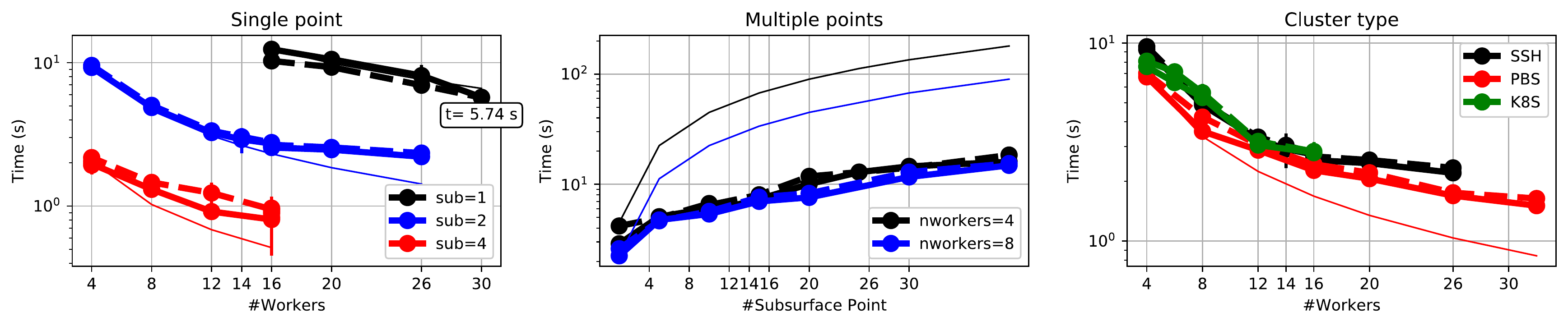}
  \caption{Benchmarking of the MDC forward (thick solid lines) and adjoint (thick dashed lines) operators with respect to a) increasing compute resources (thin lines represent the theoretical times assuming that doubling in compute resources leads to halving in compute time), b) increasing number of subsurface points (thin lines represent the time of running MDC multiple times for each subsurface point independently), c) computer cluster configurations.}
  \label{fig:benchmark}
\end{figure}

Each computation is repeated multiple times and the reported compute time is obtained as the average of each run. Figure \ref{fig:benchmark}a shows that both forward and adjoint computations have a \textit{quasi-linear} scaling with respect to increasing compute resources. This is especially the case for the Full dataset where each node performs a large enough amount of computations to saturate its hardware usage (rendering the overhead of scheduling and communication insignificant). Moreover, similar compute times for forward and adjoint passes prove that the selected strategy is optimal for both operations. In Figure \ref{fig:benchmark}b, the number of subsurface points for the input vector (i.e., $ n_{\mathbf{x}_{A}}$) is gradually increased. As matrix-matrix multiplications are performed when applying the kernel $K$ instead of matrix-vector multiplications, the compute time observed when increasing the number of points is smaller compared to the total time of running forward (or adjoint) modelling for each point separately. This property will be used in the Application section to solve the inverse problems for multiple virtual points at the same time. Finally, Figure \ref{fig:benchmark}c shows that similar performance can be achieved on different HPC systems, including the Microsoft Azure cloud environment.

\begin{table}[h!]
\centering
\begin{tabular}{ |c|c|c|c|c| } 
\hline
Geometry & Spatial sampling & Data/Kernel dimensions & Data/Kernel Size (GB) & Number of nodes \\ 
\hline
Full & 15 x 15 & 1201/300 x 9801 x 9801 & 461/230 & 16, 20, 26, 30  \\ 
\hline
Sub2 & 21 x 21 & 1201/300 x 4901 x 4901 & 115/57 & 4, 8, 12, 14, 16, 20, 26 \\ 
\hline
Sub4 & 21 x 42 & 1201/300 x 2451 x 2451 & 29/14 & 4, 8, 12, 16 \\ 
\hline
Sub8 & 21 x 84 & 1201/300 x 1126 x 1126 & 6/3 & - \\ 
\hline
\end{tabular}
\caption{Combinations of data and compute resources in the evaluation of MDC forward operator. Time domain data is of float32 type, whilst frequency domain data is of complex64 type.}
\label{table:1}
\end{table}

\section*{Applications}

\subsection{Receiver-side redatuming via Marchenko equations}
The first step of redatuming aims at moving receivers from the acquisition level to a subsurface datum of interest. By doing so, we wish to be able to handle complex propagation in the overburden and accurately estimate surface-to-subsurface Green's functions. While conventional single-scattering redatuming \cite{berryhill1984} produces inaccurate wavefields that contain spurious events arising from the incorrect handling of internal multiples in the overburden, the solution of the so-called Marchenko equations \cite{wapenaar2014} provides a means to handle all internal multiples in the redatuming process. This is accomplished by solving the following system of equations \cite{vanderneut2015}:

\begin{equation}
\label{eq:marchenko}
\begin{bmatrix}
   \Theta \mathbf{R} \mathbf{f_d^+}  \\
   \mathbf{0}
\end{bmatrix} =
\begin{bmatrix}
   \mathbf{I}  &    -\Theta\mathbf{R}   \\
    -\Theta \mathbf{R}^* & \mathbf{I}
\end{bmatrix}
\begin{bmatrix}
   \mathbf{f^-}  \\
   \mathbf{f_m^+} 
\end{bmatrix} \rightarrow \textbf{d} = \textbf{M} \textbf{f}  
\end{equation}
where $\mathbf{R}$ and $\mathbf{R}^*$ are convolution and correlation integral operators, $\Theta$ is a time-space window, $\mathbf{I}$ is the identity operator, $*$ is used to indicate complex conjugation of the kernel in frequency domain, and $\mathbf{f^-}$ and $\mathbf{f^+}=\mathbf{f}_d^+ +\mathbf{f}_m^+$ are the focusing functions to invert for. The reflection data  $R(\omega, \mathbf{x}_S, \mathbf{x}_R)$ is used as kernel for both integral operators.

Once focusing functions are estimated, the up- and down-going subsurface fields $\mathbf{g^-}$ and $\mathbf{g^+}$ can be obtained by direct evaluation of the following equation:
\begin{equation}
\label{eq:greens}
\begin{bmatrix}
   \mathbf{-g^-}  \\
   \mathbf{g^{+*}}
\end{bmatrix} =
\begin{bmatrix}
   \mathbf{I}    & -\mathbf{R}   \\
   -\mathbf{R}^* & \mathbf{I}
\end{bmatrix}
\begin{bmatrix}
   \mathbf{f^-}  \\
   \mathbf{f^+} 
\end{bmatrix}
\end{equation}
In the literature $R(\omega, \mathbf{x}_S, \mathbf{x}_R)$ is generally defined as the pressure response at $\mathbf{x}_R$ from a vertical dipole source at $\mathbf{x}_S$, with the integration in equation \ref{eq:integral} being carried out over sources. Using reciprocity, we can alternatively express the reflection response to be the vertical particle velocity recording from a monopole source; the integration in this case is carried out over the receiver axis. Mathematically speaking the two definitions are equivalent, however given the wider availability of seismic recordings with multi-component receivers, we model our data using the latter convention. The convolution integral applied by the operator $\mathbf{R}$ is therefore defined as follows:

\begin{equation}
\label{eq:integralseismic}
f^{-/+}(t, \mathbf{x}_S, \mathbf{x}_F) = \mathcal{F}_{\omega_{max}}^{-1} \left( \int_{\mathbb{S}} R(\omega, \mathbf{x}_S, \mathbf{x}_R) \mathcal{F}_{\omega_{max}}(f^{+/-}(t, \mathbf{x}_R, \mathbf{x}_F) d\mathbf{x}_R \right) 
\end{equation}

where $\mathbf{x}_F$ is the focusing point in the subsurface.

Two approaches can be taken for the solution of equation \ref{eq:marchenko}: a Neumann-type iterative scheme or inversion by means of an iterative gradient-based method. The first approach is cheaper as it generally requires less evaluations of the costly forward integral operator to converge; moreover, as it does not require the adjoint operator, its numerical implementation can be fully tailored to the forward operation. However, as discussed in \cite{dukalski2018} and \cite{zhang2019}, convergence is guaranteed only when the spectral radius of the operator $\textbf{M}$ satisfies the following condition $\rho(\textbf{M}) = \lambda_{max}(\textbf{M}) = \lim_{n\to\infty} ||\textbf{M}^n||^{1/n} < 1$. Experience has shown that this condition is hardly met in the case of highly scattering overburdens with strong contrasts, as well as when free-surface multiple reflections are included in the reflection response \cite{singh2017}. In these cases we must resolve to one of the gradient-based iteration solvers to stably solve the Marchenko equations. Whilst more demanding from a computational point of view, this approach is also beneficial in cases where the Neumann-type iterative scheme is still applicable. For example, this allows handling of missing sources in the acquisition geometry \cite{haindl2018} and a joint estimation of the focusing functions and source signature can be performed by using an alternating optimization approach \cite{becker2018}. Moreover, when dealing with ocean-bottom acquisition systems (or any system where sources and receivers are placed at different depths levels), the so-called Rayleigh-Markenko equations \cite{ravasi2016, ravasi2017, slob2013, vasconcelos2019} are required to estimate surface-to-subsurface fields. As their forward operator does not show a structure suitable for Neumann series expansion, direct inversion becomes the only viable solution.

\begin{figure}
  \includegraphics[width=\textwidth]{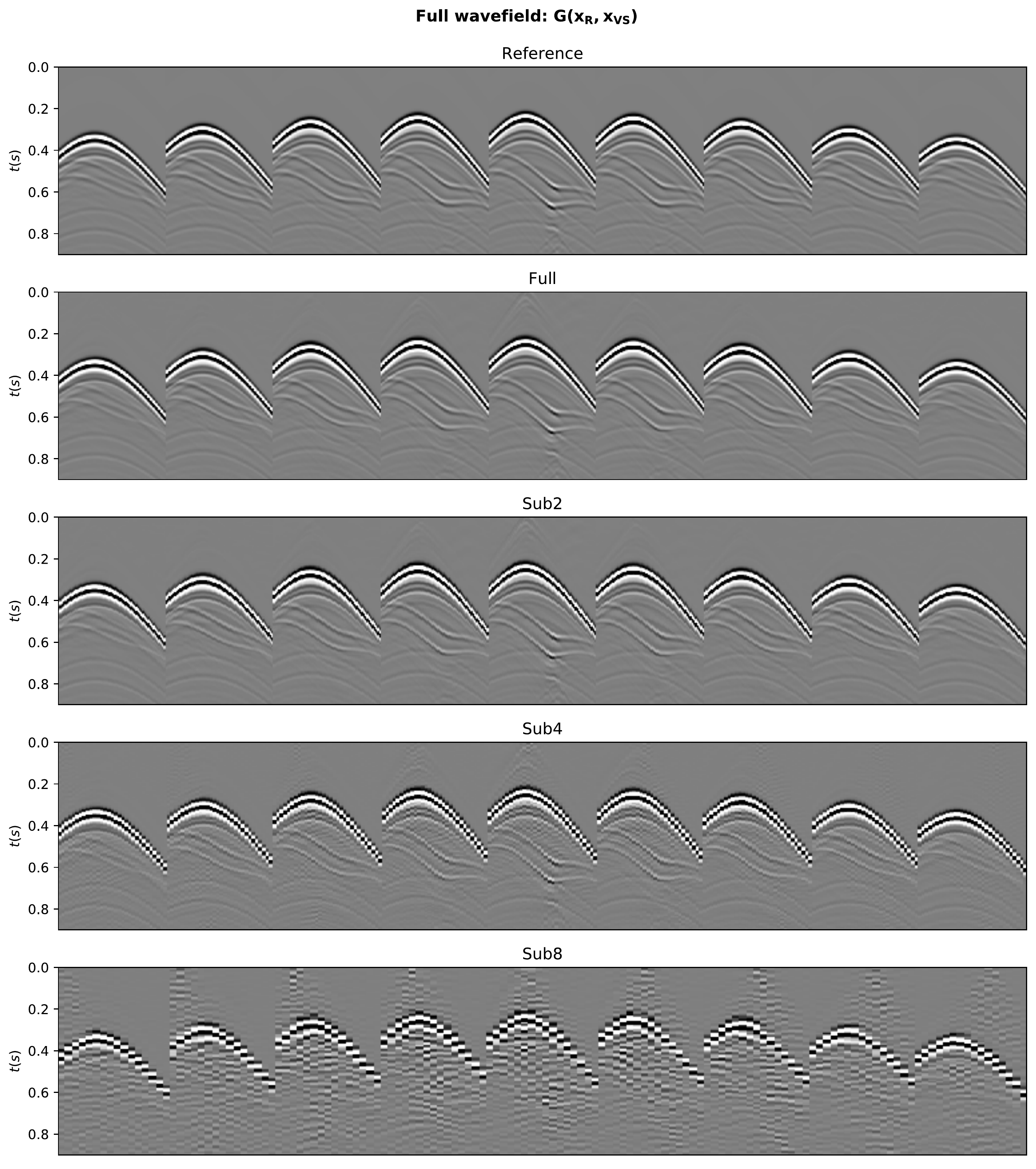}
  \caption{Full wavefield estimates. a) Reference from finite-difference modelling, and estimated by solving the Marchenko equations for b) Full data, c) Sub2 data, c) Sub4 data, and e) Sub8 data.}
  \label{fig:gest_full}
\end{figure}

\begin{figure}
  \includegraphics[width=\textwidth]{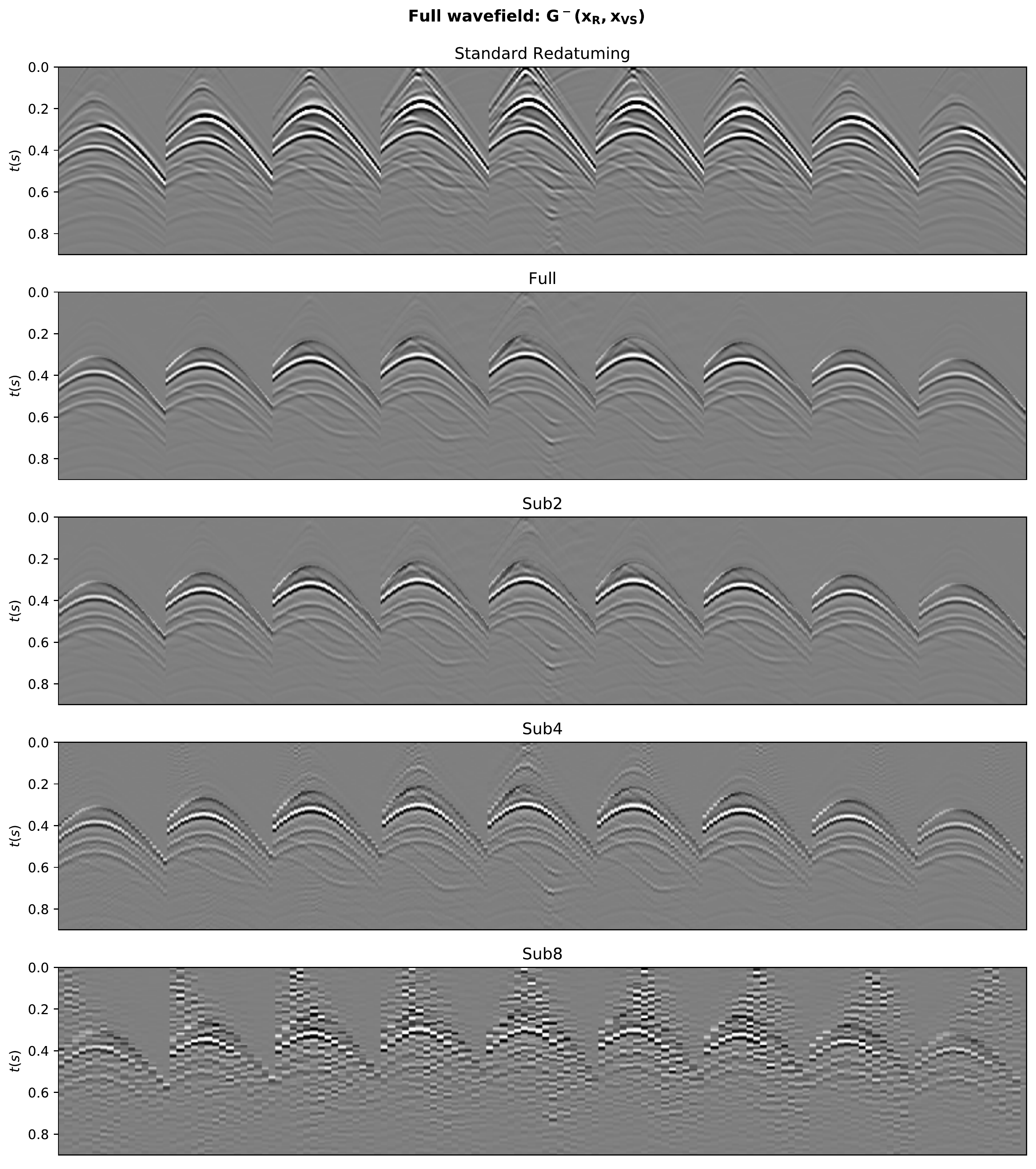}
  \caption{Up-going wavefield estimates. a) Single-scattering redatuming (i.e., using only the direct component of the downgoing focusing function in equation \ref{eq:greens}), and estimated by solving the Marchenko equations for b) Full data, c) Sub2 data, c) Sub4 data, and e) Sub8 data.}
  \label{fig:gest_up}
\end{figure}

Whilst the solution of the Marchenko equations for 3D datasets has so far been limited to the Neumann-type iterative scheme \cite{ravasi2018, jia2018, lomas2019, brackenhoff2020, staring2020}, this paper presents the first 3D example of Marchenko redatuming where the solution is obtained using a gradient-based solver. Subsurface wavefields are successfully estimated after 10 iterations of the CGLS solver (Figure \ref{fig:gest_full} and \ref{fig:gest_up}) for both the \textit{Full}, \textit{Sub2} and \textit{Sub4} data sets. Little to no aliasing effects are visible in the retrieved Green's functions and spurious events arising from incorrect handling of overburden propagation in single-scattering redatuming (bottom left panel in Figure \ref{fig:gest_up} -- $\mathbf{g_0^-} = \mathbf{R} \mathbf{f_d^+}$) are successfully suppressed. Note that the source/receiver spacing for the \textit{Sub4} dataset is greater than a quarter of the dominant wavelength ($\lambda_{dom} / 4 = c / (4*f_{dom}) = 30 m$) in the coarser direction; this result suggests that the ability to remove artefacts from the up-going wavefield may be less affected by coarse spatial sampling in three-dimensions than in its two-dimensional counterpart (Peng and Vasconcelos, 2019; Lomas and Curtis, 2019). Strong aliasing does however arise when solving the Marchenko equations with an even coarser acquisition geometry (i.e., \textit{Sub8} dataset) showing that strict acquisition requirements do still apply to the 3D case.

\begin{figure}
  \includegraphics[width=\textwidth]{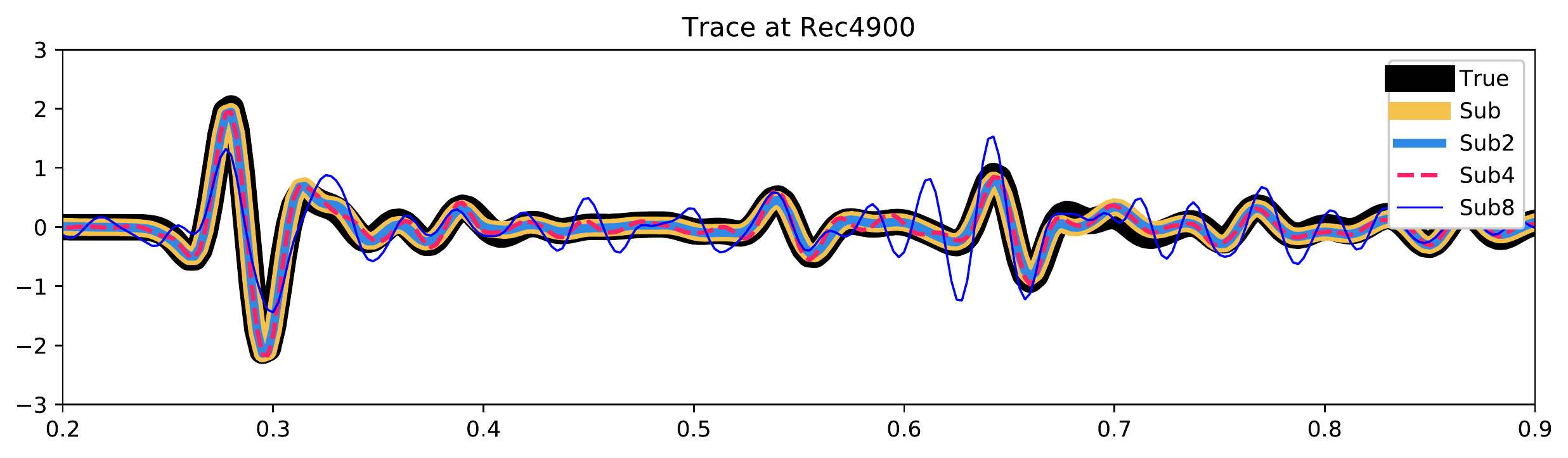}
  \caption{Trace comparison of the full wavefield for Receiver 4900 ($e^{3t}$ applied to traces).}
  \label{fig:gest_trace}
\end{figure}

\subsection{Source-side redatuming via Multi-dimensional deconvolution}
As observed in Figures \ref{fig:gest_full} and \ref{fig:gest_up}, the retrieved Green's functions with subsampling factors of 2 and 4 show minimal degradation in terms of quality and aliasing effects. On the other hand, the responses retrieved using the \textit{Sub8} dataset show a stronger degree of noise due to poor sampling of the spatial integration axis in equation \ref{eq:integralseismic}. Nevertheless, our ability to correctly solve the Marchenko equations and remove spurious events present in the standard redatumed up-going Green's function (top panel in Figure \ref{fig:gest_up}) does not seem to be greatly affected. In order to assess the validity of this observation, a second step of redatuming is required. 

By moving sources to the same depth level as the previously redatumed receivers, we create the so-called local reflectivity of the subsurface in the target area below the focusing datum. Further to that, we can assess the strength of spurious reflectors in both virtual common-shot gathers and zero-offset sections. More specifically, provided the availability of full-wavefield up- ($g^-$) and down-going ($g^+$) band-limited Green's functions, the local reflectivity $r(t, \mathbf{x}_F, \mathbf{x}_{F'})$ can be estimated by solving the so-called Multi-dimensional Deconvolution (MDD) equations:
\begin{equation}
\label{eq:mdd}
g^{-}(t, \mathbf{x}_{S}, \mathbf{x}_{F'}) = \mathcal{F}_{\omega_{max}}^{-1} \left( \int_{\mathbb{S}} G^+(\omega, \mathbf{x}_S, \mathbf{x}_F) \mathcal{F}_{\omega_{max}}(r(t, \mathbf{x}_F, \mathbf{x}_{F'}) d\mathbf{x}_F \right)  \quad 
\end{equation}
which can be discretized and expressed in compact matrix-vector notation as follows, $\mathbf{g}^- = \mathbf{G}^+ \mathbf{r}$. This equation represents a Fredholm integral of the first kind that cannot be solved via Neumann series expansion. The same iterative solver employed in the previous example is be used instead. 

Computing the local reflectivity response by simply applying the adjoint step of equation \ref{eq:mdd} is equivalent to both the well-known interferometric-based redatuming method of \cite{bakulin2006} and the correlation-based imaging condition, commonly used by wave-equation based imaging algorithms such as reverse-time migration; as a consequence of not solving the underlying inverse problem, source-side propagation in the overburden is not properly handled and artefacts appear in the estimated reflectivity response. Even when the input Green's functions do not contain any spurious arrivals from the receiver-side redatuming step, artefacts appear due to cross-talk between unrelated events in the up- and down-going components. 

An alternative approach that can alleviate this problem has been recently proposed by \cite{staring}. The so-called double-focusing method replaces the full up- and down-going Green's functions by the first two terms of the Neumann series expansion of the Marchenko equations. These terms are then adaptively summed leading to the cancellation of all purely overburden-related artefacts. Whist not requiring any inversion and being robust because of its adaptive nature, this method does not use the entire knowledge of the coda in the downgoing wavefield. As a result, the amplitude-variation-with-offset behaviour of the different events in the retrieved local reflectivity is still affected by illumination unbalancing which can only be corrected by fully solving equation \ref{eq:mdd} \cite{ravasi2017}. 

In this example, we estimate up- and down-going Green's functions by solving the Marchenko equations for a grid of points at depth of $z_F = 650 m$. The grid is composed of 2911 points, sampled from $x_{F,in} = 200 m$ to $x_{F,end} = 1600 m$ with spacing $dx_F = 20 m$ along the x axis, and from $y_{F,in} = 200 m$ to $y_{F,end} = 1000 m$ with spacing $dy_F = 20 m$ along the y axis. Equation \ref{eq:mdd} is then repeatedly solved for batches of 20 virtual sources $\mathbf{x}_{F'}$ at a time. Whilst solving the entire problem at once (i.e., with all virtual sources) brings some additional benefits, such as the possibility to add reciprocity constraints \cite{vanderneut2017}, it becomes prohibitive for problems of this size. The model ($r$) and data ($g^-$) vectors do in fact dramatically increase in size, violating our third assumption in the implementation of the batch matrix multiplication operator. This does in turn lead to a high communication overhead at every pass of MDC (and its adjoint) and poor performance.

\begin{figure}
  \centering
  \includegraphics[width=0.7\textwidth]{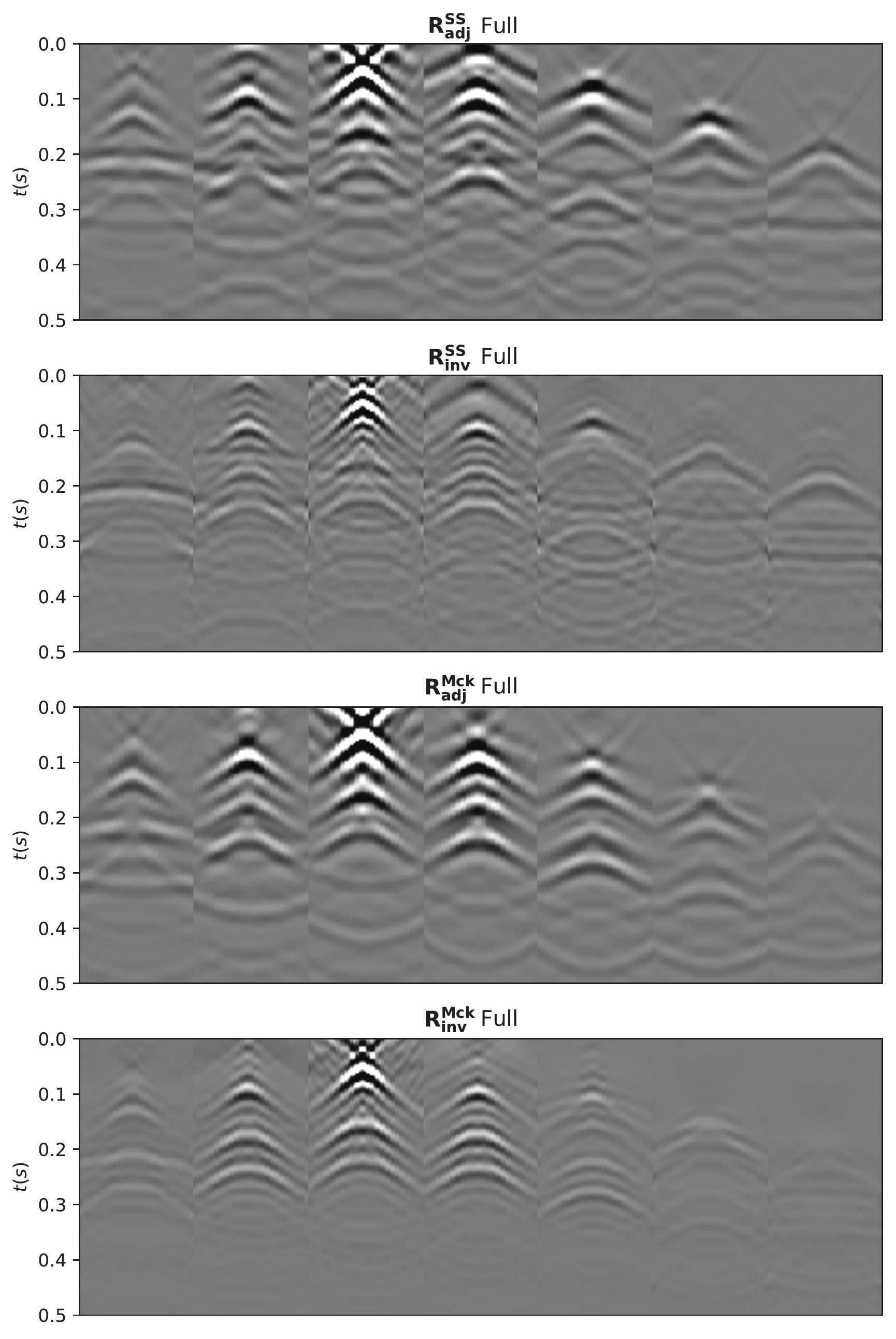}
  \caption{Local reflectivites. Reflectivities are estimated using a) single-scattering redatumed fields and the adjoint of equation \ref{eq:mdd}, single-scattering redatumed fields and the inverse of equation \ref{eq:mdd}, a) Marchenko redatumed fields and the adjoint of equation \ref{eq:mdd}, Marchenko redatumed fields and the inverse of equation \ref{eq:mdd}.}
  \label{fig:localrefl}
\end{figure}

Figure \ref{fig:localrefl} shows a portion of the retrieved local reflectivity for a virtual source in the redatuming grid (yellow dot in Figure \ref{fig:geom}) and several lines of virtual receivers (cyan lines in Figure \ref{fig:geom}). The top two panels have been created by applying the adjoint and inverse of equation \ref{eq:mdd} to the Green's functions estimated via single-scattering redatuming (i.e., by plugging the initial focusing function $\textbf{f}^+_d$ into equation \ref{eq:greens}). The Green's functions estimated by solving the Marchenko equations are instead used to produce the reflectivities in the two bottom panels. Inversion of equation \ref{eq:mdd} alone is not able to correct for errors in the input datasets and leads to strong artefacts as visible in the second panel. On the other hand, by using accurate Green's functions and inverting equation \ref{eq:mdd}, we can produce very satisfactory estimates of the local reflectivity. Such reflectivity contains 3 main events arising from the corresponding discontinuities in the density model below the focusing datum. The lack of full illumination at the edges of the grid leads to degradation of the estimated local reflectivity as distance from the virtual source increases. This amplitude behaviour is also partially due to the fact that the local reflectivity is a particle velocity response from a monopole source; it does thus carry the directional radiation pattern of a velocity field. Finally, a clear difference is observed in the frequency content of the reflectivities obtained from the adjoint and inverse operators; this is due to the fact that wavelet deconvolution is also achieved when solving equation \ref{eq:mdd}.

\begin{figure}
  \centering
  \includegraphics[width=0.9\textwidth]{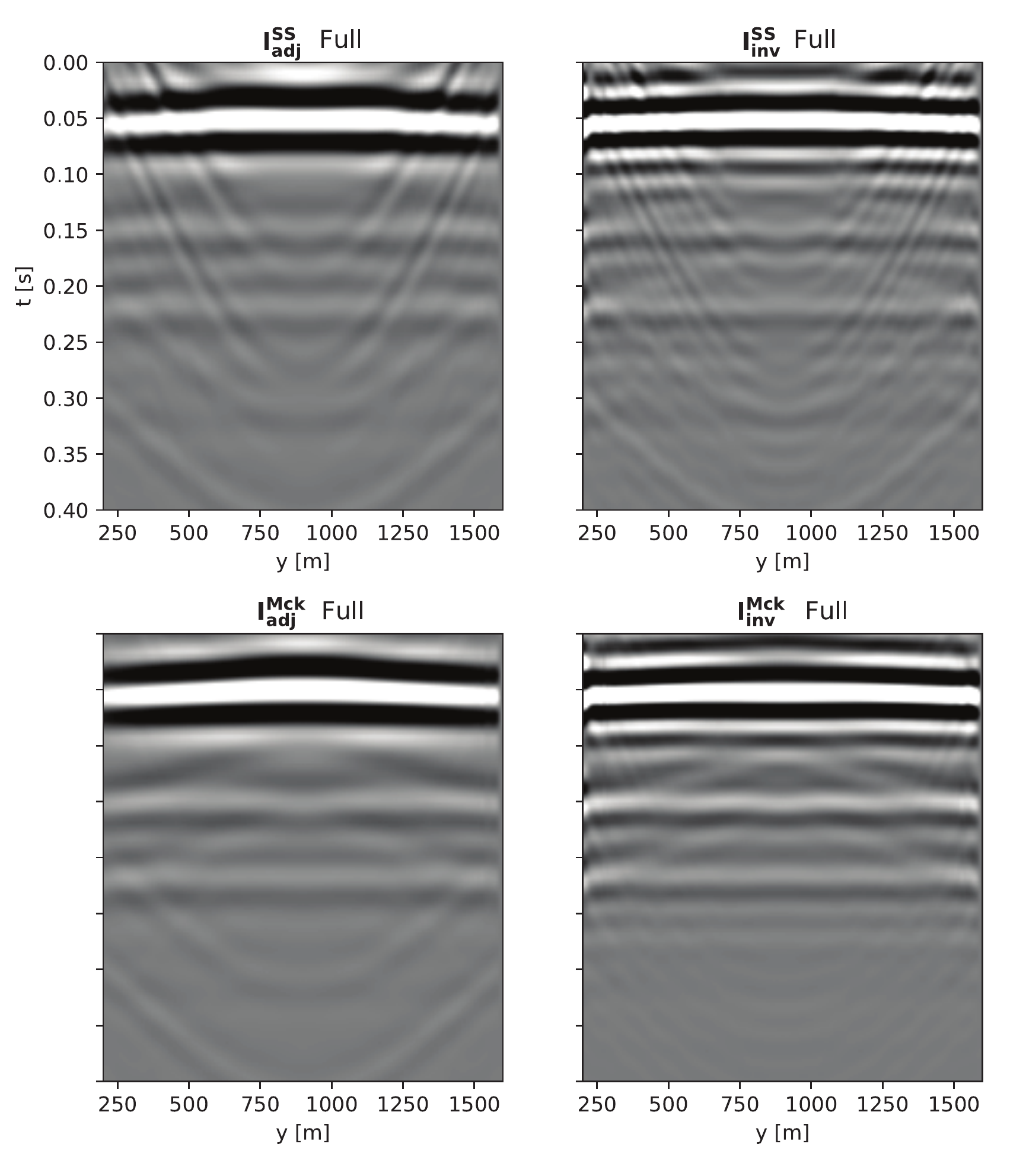}
  \caption{Zero-offset sections extracted from the local reflectivities in Figure \ref{fig:localrefl}, ordered in the same fashion as in the previous figure.}
  \label{fig:images}
\end{figure}

Figure \ref{fig:images} shows the corresponding zero-offset sections along the middle cyan line in Figure \ref{fig:geom}, which can be interpreted as local \textit{time} images of the subsurface beneath the focusing datum. Similarly to the local reflectivities, artefacts are partially removed in the third panel and further suppressed in the fourth panel, which is obtained from the inverted local reflectivity using the Marchenko Green's functions as input.

As a final remark, we note that to take full advantage of all the available offsets, the estimated local reflectivity could be used as input to target-oriented imaging and inversion algorithms: examples of such a kind are localized imaging via RTM \cite{wapenaar2014, ravasi2017}, localized FWI \cite{cui2018} or localized 1D inversion with multiply scattered waves \cite{gisolf2017}. In all of the above scenarios, the ability to redatum an entire acquisition level closer to a target of interest without compromising on the data quality allows for the introduction of more complex physics (e.g., elastic wave equation) as well as for increasing the resolution of the estimated model parameters, making those products suitable for subsequent reservoir characterization tasks. 

\section*{Discussion}

\subsection*{Solving MDD using time-domain multi-dimensional convolution operators}
Historically, multi-dimensional deconvolution problems like those in equation \ref{eq:mdd} have been solved in the frequency domain. Since each frequency component can be computed and inverted individually, this approach is computationally attractive and it also allows for the adoption of very simple parallelization strategies with no exchange of data between each subproblem. However, the ill-posedness of the problem calls for the inclusion of prior information (i.e., regularization and/or preconditioning) to improve the stability of the overall inverse process \cite{minato}. As frequencies show different levels of ill-posedness, with low frequencies being generally more stable then high frequencies, different regularizations have to be carefully selected when solving each subproblem independently. Failing to do so results in incorrectly balanced frequency components in the estimated solution, which may ultimately lead to additional noise in its time domain counterpart.

\cite{vanderneut2015b, vanderneut2017} suggested to tackle the MDD problem directly in the time-space domain. Whilst this alternative approach requires to solve a single, much larger, system of equations instead of multiple independent subproblems, it carries some important advantages over the former approach. First, it relieves from the need to defining frequency-dependant regularization parameters. Second, additional time-domain preconditioners (e.g., causal time windowing) can be added to aid the inversion. In this paper, we have shown that by solving equation \ref{eq:mdd} with an iterative solver no additional regularization is required to obtain a stable solution even when the the input wavefields are not directly recorded data, rather the product of a precedent inverse problem. We conjecture that the use of an iterative solver and a stopping criterion based on the residual norm acts as a  natural regularizer to the problem as it prevents for entirely fitting noise in the data term. This further explains the value of using MDC operators in the time-domain also in problems where a solution in the frequency domain is theoretically possible.

\subsection*{Computational cost of MDC-based inversion workflows}
In this section, we carry out an analysis of the overall computational cost involved in the solution of the two inverse problems presented in the Application section. In order to keep the section concise, we assume that the cost of solving such inverse problems is dominated by the number of convolutional integrals that are performed throughout the iterations of the chosen solver (in our case CGLS).

Starting from equation \ref{eq:marchenko}, the overall modelling operator for the Marchenko equations is a $2 \times 2$ block matrix which contains two multi-dimensional convolution operators, namely $\textbf{R}$ and $\mathbf{R}^*$. The cost of performing the forward and adjoint steps is therefore quantified as 2 MDC operations. As each iteration of CGLS requires the application of one forward one and adjoint passes, the overall cost of one iteration of CGLS is 4 MDC operations. Following our numerical example where that acceptable convergence is reached after $N_{iter}=10$ iterations of CGLS, the overall cost of the inverse problem is 40 MDC operations. Noting that an additional MDC operation is required to create the data vector in equation \ref{eq:marchenko} and that two additional MDC operations are required to compute the Green's functions (equation \ref{eq:greens}), the overall cost of estimating subsurface-to-surface fields is in the order of 43 MDC operations. 
A similar analysis can be carried out for the multi-dimensional deconvolution problem in equation \ref{eq:mdd}. In this case the data vector is already available and the modelling operator is directly represented by a multi-dimensional convolution operator with the downgoing Green's function. In the solution of the inverse problem, it is therefore only required to perform $N_{iter}$ MDC operations. 

Finally, as the MDC operator can be applied to multiple virtual points at the same time (Figure \ref{fig:benchmark}b), we can also solve equations \ref{eq:marchenko} and \ref{eq:mdd} for a certain number of focusing points ($N_F$) instead of solving $N_F$ independent system of equations in parallel. This number should be chosen as large as possible whilst ensuring that the communication overhead associated to every MDC step does not become too large and dominates over the advantage of performing one matrix-matrix multiplication over many matrix-vector multiplications. However, as the number of focusing points for which we are interested to perform redatuming is generally on the same order of magnitude (or possibly even larger) than the number of sources and receivers, multiple groups of focusing points are generally created and inverted separately. 

The overall redatuming process can therefore take advantage of 3 levels of parallelism:

\begin{itemize}
  \item Embarrassingly parallel parallelism: the entire redatuming job is divided into a number of independent tasks, each of them comprising of an inversion for a group focusing points;
  \item Distributed parallelism: each independent task is performed using a group of Dask workers (alongside a Dask scheduler and client) which have enough memory to load the entire convolutional kernel in memory. Each task solves multiple inverse problems sequentially; 
  \item Multi-core parallelism: each worker performs several operations that can take advantage of highly efficient NumPy routines (e.g., \textit{np.matmul} for batched matrix multiplication) that leverage multithreading within the compute node.
\end{itemize}

Table \ref{table:2} summarizes the timings related to the loading of the input dataset and setting up of the inverse problem, as well as those related to the solution of the inverse problem itself for the different subsampled data. 

\subsection*{Cloud computing}
Up until now we have not discussed in details the HPC architecture used to run our computations. This is because, by leveraging the Dask library, we can easily deploy our solution onto any HPC clusters that implements one of the queuing systems supported by Dask. Figure \ref{fig:benchmark}c shows that we can achieve similar performance independently on the HPC solution adopted.

Nevertheless, the ability to perform parallelism at different levels makes \textit{cloud computing} a cost effective alternative to on-premise computing for such type of workloads. This comes with the obvious benefit of providing theoretically unlimited scalability without any up-front cost. A particularly appealing feature of most cloud computing environments is represented by the so-called \textit{spot pricing}, which provides access to unused compute capacity at deep discount. However, as the underlying compute resource can be evicted at any time if the cloud provider needs capacity, only workloads that can sustain and/or recover from interruptions are suited to run on spot instances. This feature is of particular interest for both the Marchenko redatuming and MDD workflows as we generally solve the same inverse problem many times for a single (or groups of) independent subsurface points. Whilst an upfront time is required for new resources to boot-strap and start running computations, each inverse problem is then completely independent from the others, meaning that the only wasted computation if a node is evicted before the inverse process is terminated is that of the initial setup and the current inversion that was running. Our experience shows that eviction does not happen very often, and we can thus obtain a great cost reduction when running these workflows with spot instances compared to guaranteed resources (Table \ref{table:3}). 

By leveraging the orchestration capabilities offered by Kubernetes (K8S), limited effort is required to adapt codes that were developed for classic HPC environments to be run on cloud environments. More specifically, starting from the freely available Dask Helm charts\footnote{https://github.com/dask/helm-chart}, we have created a new chart that contains amaster pod running a K8S \textit{Job}. This pod is responsible for setting up the inverse problem of interest for a number of subsurface points and keeping track of the number of inversions that have been carried out to completion. Another K8S pod, physically living on the same node (or another), is instantiated for the Dask scheduler and a number of other Pods for the Dask workers. Any time a node is evicted and the corresponding pod(s) fail, K8S is in charge of automatically re-instatiating the required resources. If any of the Worker is permanently deleted, Dask will handle this event by redistributing the lost portion of the kernel to the other Worker nodes and the main job is preserved. On the other hand, if either the Scheduler or Master pods are deleted, an entire new job is restarted from the subsurface point at which failure was experienced. As K8S is responsible to ensure that a job is successfully terminated, our responsibility is limited to keeping track of the evolution of our inversion and ensure that the new Job restarts the computation from where it had stopped. 

\begin{table}[h!]
\centering
\begin{tabular}{ |c|c|c|c|c| } 
\hline
Geometry & Nodes & Preparation (s) & Inversion of 1 point (s) & Inversion of 20 points (s) \\ 
\hline
Full & 16 & 180 & 260 & 1760 \\ 
\hline
Sub2 & 8 & 55 & 115 & 930 \\ 
\hline
Sub4 & 8 & 30 & 44 & 460 \\ 
\hline
\end{tabular}
\caption{Timings of the different steps involved in the solution of the Marchenko equations as an inverse problem for one or multiple focusing points. Intel(R) Xeon(R) CPUs @ 2.90GHz with 16 threads and 128GB of RAM each are used for these computations.}
\label{table:2}
\end{table}

\begin{table}[h!]
\centering
\begin{tabular}{|l|l|l|l|l|}
\hline
                             & On-demand & Spot ($\sim3h$ eviction) & Spot ($\sim6h$) & Spot ($\sim12h$) \\ \hline
Pod(s) creation (min)        & $\sim$ 3    & 160                      & 80              & 40               \\ \hline
Setup (min)                  & 3         & 240                      & 120             & 60               \\ \hline
N. of total evictions        & 0         & 80                       & 40              & 20               \\ \hline
Proc. time (h) & 242.58      & 307.27                               & 274.93          & 258.76           \\ \hline
Total time (h)               & 242.67    & 313.93                   & 278.26          & 260.42           \\ \hline
Projected Cost (\$)          & 3913.72   & 1199.50                  & 1063.18         & 995.02           \\ \hline
\end{tabular}
\caption{Costs associated with the entire receiver-side redatuming process for the \textit{Sub1} dataset ($N_F=2911$ subsurface points) in a cloud environment using both on-demand and spot resources. These costs are computed using the hourly price of a \textit{Standard\_E16\_v3} instance at the time of writing (1,008 \$/h) as well as the hourly price that we were able to obtain for the same instance in a SPOT configuration (0,2388 \$/h)}
\label{table:3}
\end{table}

\subsection*{Spatial sampling and aliasing effects on imaging products}
\begin{figure}
  \centering
  \includegraphics[width=0.9\textwidth]{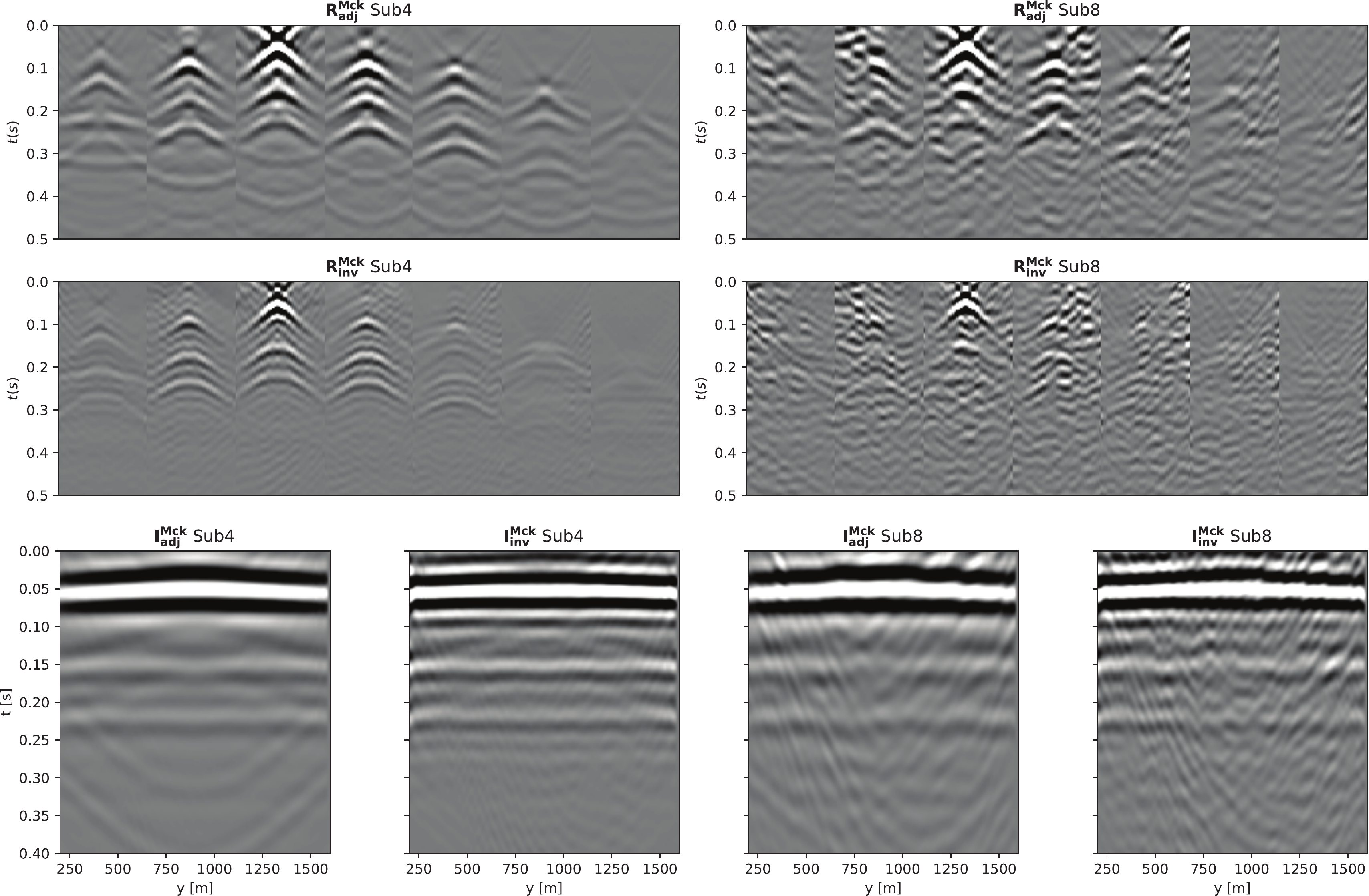}
  \caption{Local reflectivities from the Marchenko redatumed fields using a-b) the adjoint and c-d) the inverse of equation \ref{eq:mdd} for the \textit{Sub4} and \textit{Sub8}, respectively. e-f-g-h) Corresponding zero-offset sections.}
  \label{fig:imagessub}
\end{figure}

In this last section we first analyse the effect of regular subsampling on the source-side redatuming step. The up- and down-going Green's functions, estimated by solving the Marchenko equation at a constant depth level ($z_F = 650 m$), are now used as an input to equation \ref{eq:mdd} and local reflectivities are retrieved by means of least-squares inversion with the same number of iterations used to invert the \textit{Full} dataset. As the results for the \textit{Sub2} dataset are practically identical to those from the \textit{Full} dataset, we avoid showing them and focus on the other two subsampled versions of the original dataset. In Figure \ref{fig:imagessub} the local reflectivities from a virtual source in the middle of the redatuming grid and zero-offset images along a single crossline are shown for both datasets. It is possible to notice how by subsampling both the receiver and source arrays by a factor of 4, no severe degradation in the source-side redatuming step is observed apart from the introduction of some small coherent noise in the far offsets. This is possibly due to a poor sampling of some of the stationary contributions in the spatial integral in equation \ref{eq:mdd}. On the other hand, when applying a subsampling factor of 8, the retrieved reflectivities are of noticeably poorer quality. This is especially the case when we attempt to invert equation \ref{eq:mdd}, while correlation-based redatuming shows less sensitivity to a coarsening of the spatial integration axis. Similarly, the images for the \textit{Sub4} dataset are of very similar quality to those of the \textit{Full} dataset, both for correlation-based and inversion-based source-side redatuming. On the other hand, whilst the key reflectors are still visible in the images from the Sub8 dataset, we can conclude that by heavily subsampling the acquisition geometry, the resulting images are affected by strong coherent noise, which becomes even more prominent when attempting to image the up- and down-going Green's functions by means of inversion. Whilst the correlation-based imaging condition appears to be less sensitive to poor spatial sampling, migration artefacts due to cross-talk between unrelated events in the up- and down-going Green's function do however show up as expected from our previous imaging experiments.

\begin{figure}
  \includegraphics[width=\textwidth]{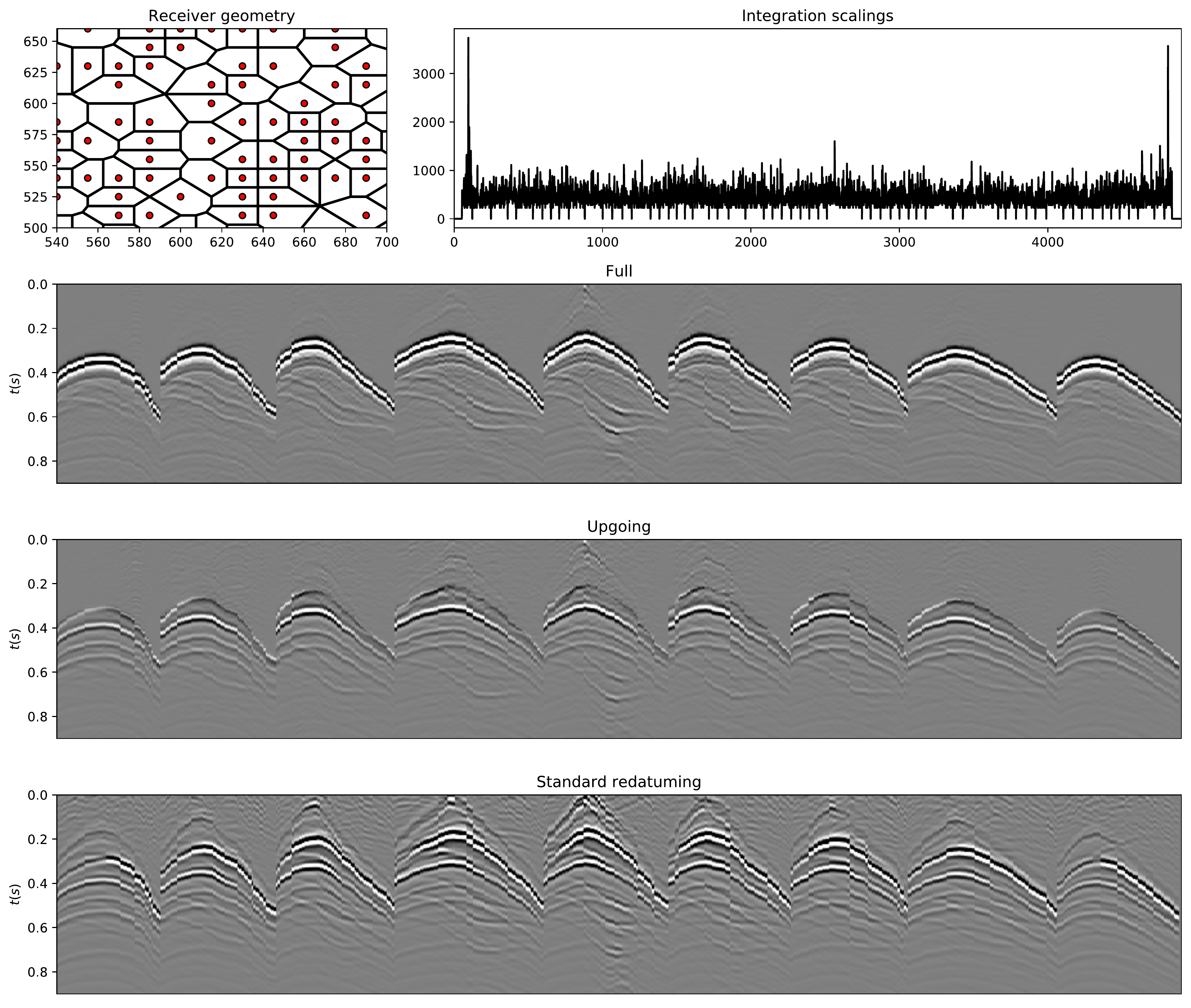}
      \caption{Marchenko redatuming for a irregularly sampled acquisition geometry. a) Close-up of receiver (and source) distribution with Voronoi tesselation. b) Integration scaling factors used for each receiver based of the area of its Voronoi cell. c) Full, b) up-going, and c) single-scattering wavefields obtained by solving the Marchenko equations.}
  \label{fig:gest_irreg}
\end{figure}

Finally, we also consider the case of irregular sampling. The acquisition geometry is now created by randomly selecting half of the original set of sources and receivers, which were regularly sampled over a rectangular grid with spatial sampling of 15 both in the x and y axes. Note that the number of traces in this dataset are thus the same as those in the Sub2 dataset. A pragmatic approach to handle such a spatial irregularity consists of performing Voronoi tessellation of the available receivers to determine the contribution that each receiver has in the spatial integration in equation \ref{eq:integralseismic}, which is proportional to the areal extent of the corresponding voxel (Figure \ref{fig:gest_irreg}a) . Such receiver-dependant scaling (Figure \ref{fig:gest_irreg}b) can be applied to each shot gather as part of the pre-processing of the reflection response. The inverse problem in equation \ref{eq:marchenko} is then solved using CGLS and the same number of iterations as in the other examples and the retrieved Green's functions are shown in Figure \ref{fig:gest_irreg}c and d. Apart from the jittering effect in the seismic events in the retrieved Green's functions, arising from the fact that sources are irregularly sampled, this example shows that by carefully handling the contribution of each receiver in the spatial integration step in equation \ref{eq:integralseismic}, the Marchenko equations can be satisfactorily inverted also in the presence of completely irregular geometries. This would have not been the case if we have very sparse sources (and receivers) as this would ultimately lead to aliasing effects when performing spatial integration similar to those observed for the \textit{Sub8} dataset. Moreover, this example still assumes co-location between sources and receivers in the irregular grid, which is never the case in real life applications. In the case when the availability of sources if limited to only a portion of the receiver locations, the modified formulation of \cite{haindl2018} could be used to deal with missing sources. Although applied to 2D datasets, its extension to the three dimensions is trivial. Alternatively, when receivers are heavily subsampled leading to a coarse and irregular spatial integration support, the resulting fields will be a blurred version of the equivalent fields obtained from a finely sampled grid; a modification of the Neumann-like iterative scheme has been recently proposed by \cite{wapenaar2020} to handle such a scenario.

\section*{Conclusion}
In this work we have presented a framework for the computation of out-of-core multi-dimensional convolution integral operators and discussed how they can be efficiently implemented on distributed computer systems. The proposed solution is shown to handle kernel operators whose size exceeds hundreds of Gigabytes and to scale with available compute resources. Furthermore, by providing implementations of the forward and adjoint passes with similar performances, it allows for solving inverse problems that require repeated application of such operators. To illustrate its applicability, an example of full-wavefield redatuming for a synthetic 3D seismic dataset is presented. Being able to achieve such a goal open doors to a large variety of novel target-oriented imaging and inversion applications to be successfully applied to 3D datasets acquired in real life scenarios.

\section*{Acknowledgements}
The authors thank their respective institutions for allowing to present this work. MR thanks Haithem Jarraya and Stian Øvrevåge for the support with K8S. The different components of the computational framework presented in the work are freely available and can be installed using standard distribution systems such as PyPI and conda-forge. We refer the reader to the following Github repository (https://github.com/mrava87/EAGE\_MDCHPC\_2020), which contains a set of Jupyter notebooks and Python scripts used to produce the figures in this manuscript as well as the Helm charts for running our workflows on a Kubernetes cluster.

\bibliographystyle{unsrt}  
\bibliography{references}

\begin{thebibliography}{10}

\bibitem{dragoset2010}
B.~Dragoset, E.~Verschuur, I.~Moore, and R.~Bisley.
\newblock A perspective on 3d surface-related multiple elimination.
\newblock {\em Geophysics}, 75:75A245--75A261, 2010.

\bibitem{pica2005}
A.~L. Pica, G.~Poulain, B.~David, M.~Magesan, S.~Baldock, T.~Weisser,
  P.~Hugonnet, and P.~H. Herrmann.
\newblock 3d surface-related multiple modeling, principles and results.
\newblock In {\em 75th Annual International Meeting, SEG, Expanded Abstracts},
  pages 2080--2083, 2005.

\bibitem{vangroenestijn2009}
G.~J. van Groenestijn and D.~J. Verschuur.
\newblock Estimating primaries by sparse inversion and application to
  near-offset data reconstruction.
\newblock {\em Geophysics}, 74:1MJ--Z54, 2009.

\bibitem{lopez2015a}
G.~A. Lopez and D.J. Verschuur.
\newblock Closed-loop surface-related multiple elimination and its application
  to simultaneous data reconstruction.
\newblock {\em Geophysics}, 80, 2015.

\bibitem{lin2013}
T.~Y. Lin and F.~J. Herrmann.
\newblock Robust signature deconvolution and the estimation of primaries by
  sparse inversion.
\newblock {\em Geophysics}, 78(3):R133–R150, 2013.

\bibitem{lopez2015b}
G.~A. Lopez and D.J. Verschuur.
\newblock 3d focal closed-loop srme for shallow water.
\newblock {\em Geophysical Journal International}, 203:792--813, 2015.

\bibitem{amundsen2001a}
L.~Amundsen.
\newblock Elimination of free-surface related multiples without need of a
  source wavelet.
\newblock {\em Geophysics}, 66:327--341, 2001.

\bibitem{ravasi2015}
M.~Ravasi, I.~Vasconcelos, A.~Curtis, and A.~Kritski.
\newblock Multi-dimensional free-surface multiple elimination and source
  deblending of volve obc data.
\newblock {\em 77th EAGE Annual International Conference and Exhibition}, 2015.

\bibitem{vanderneut2015}
J.~van~der Neut, I.~Vasconcelos, and K.~Wapenaar.
\newblock On green's function retrieval by iterative substitution of the
  coupled marchenko equations.
\newblock {\em Geophysical Journal International}, 203:792--813, 2015.

\bibitem{dukalski2018}
M.~Dukalski and K.~de~Vos.
\newblock Marchenko inversion in a strong scattering regime including
  surface-related multiples.
\newblock {\em Geophysical Journal International}, 212:760–776, 2018.

\bibitem{becker2018}
T.~S. Becker, M.~Ravasi, D.J. van Manen, F.~Broggini, and J.O.A. Robertsson.
\newblock Sparse inversion of the coupled marchenko equations for simultaneous
  source wavelet and focusing functions estimation.
\newblock {\em 80th EAGE Annual International Conference and Exhibition}, 2018.

\bibitem{ravasi2017}
M.~Ravasi.
\newblock Rayleigh-marchenko redatuming for target-oriented, true-amplitude
  imaging.
\newblock {\em Geophysics}, 82(6):S439--S452, 2017.

\bibitem{vasconcelos2019}
I.~Vasconcelos and Y.~Sripanich.
\newblock Scattering-based marchenko for subsurface focusing and redatuming in
  highly complex media.
\newblock {\em 81st EAGE Annual International Conference and Exhibition}, 2019.

\bibitem{zhang2019}
L.~Zhang, J.~Thorbecke, K.~Wapenaar, and E.~Slob.
\newblock Transmission compensated primary reflection retrieval in the data
  domain and consequences for imaging.
\newblock {\em Geophysics}, 84(4):Q27--Q36, 2019.

\bibitem{haindl2018}
C.M. Haindl, F.~Broggini, M.~Ravasi, and D.J. van Manen.
\newblock Using sparsity to improve the accuracy of marchenko imaging given
  imperfect acquisition geometries.
\newblock {\em 80th EAGE Annual International Conference and Exhibition}, 2018.

\bibitem{wapenaar2010}
K.~Wapenaar and J.~van~der Neut.
\newblock A representation for green's function retrieval by multidimensional
  deconvolution.
\newblock {\em JASA}, 128:EL366--EL371, 2010.

\bibitem{wapenaar2011}
K.~Wapenaar, J.~van~der Neut, E.~Ruigrok, D.~Draganov, J.~Hunziker, E.~Slob,
  J.~Thorbecke, and R.~Snieder.
\newblock Seismic interferometry by crosscorrelation and by multidimensional
  deconvolution: A systematic comparison.
\newblock {\em Geophysical Journal International}, 185:1335--1364, 2011.

\bibitem{vanderNeut2013}
J.~van~der Neut and F.~Herrmann.
\newblock Interferometric redatuming by sparse inversion.
\newblock {\em Geophysical Journal International}, 192:666--670, 2013.

\bibitem{vanderneut2017}
J.~van~der Neut, M.~Ravasi, Y.~Liu, and I.~Vasconcelos.
\newblock Target-enclosed seismic imaging.
\newblock {\em Geophysics}, 82(6):Q53--Q66, 2017.

\bibitem{vasconcelos2017}
I.~Vasconcelos, M.~Ravasi, A.~Kritski, J.~van~der Neut, and T.~Cui.
\newblock Local, reservoir-only reflection and transmission responses by
  target-enclosing extended imaging.
\newblock {\em SEG Technical Program Expanded Abstracts}, pages 5289--5293,
  2017.

\bibitem{berkhout2011}
A.J Berkhout and D.J. Verschuur.
\newblock Full wavefield migration, utilizing surface and internal multiple
  scattering.
\newblock In {\em 81st Annual International Meeting, SEG, Expanded Abstracts},
  pages 3212--3216, 2011.

\bibitem{davydenko2016}
M.~Davydenko and E.~Verschuur.
\newblock Full-wavefield migration: using surface and internal multiples in
  imaging.
\newblock {\em Geophysical Prospecting}, 65, 2016.

\bibitem{staal2013}
X.R. Staal and D.J. Verschuur.
\newblock Joint migration inversion, imaging including all multiples with
  automatic velocity update.
\newblock {\em 75th EAGE Annual International Conference and Exhibition}, 2013.

\bibitem{bisley2005}
R.~Bisley, I.~Moore, and H.~W. Dragoset.
\newblock Generalized 3d surface multiple prediction, 2005.

\bibitem{Wapenaar2014a}
K.~Wapenaar, J.~Thorbecke, J.~van~der Neut, F.~Broggini, E.~Slob, , and
  R.~Snieder.
\newblock Marchenko imaging.
\newblock {\em Geophysics}, 79:WA39--WA57, 2014.

\bibitem{jia2018}
X.~Jia, Y.~Zhao, and R.~Snieder.
\newblock Data interpolation for 3d marchenko green’s function retrieval.
\newblock {\em SEG Technical Program Expanded Abstracts}, page 4588–4592,
  2018.

\bibitem{lomas2019}
A.~Lomas and A.~Curtis.
\newblock Marchenko methods in a 3-d world.
\newblock {\em Geophysical Journal International}, 220:296–307, 2019.

\bibitem{brackenhoff2020}
J.~Brackenhoff, J.~Thorbecke, V.~Koehne, D.~Barrera, and K.~Wapenaar.
\newblock Implementation of the 3d marchenko method, 2020.

\bibitem{berryhill1984}
J.~R. Berryhill.
\newblock Wave-equation datuming before stack.
\newblock {\em Geophysics}, 49:2064–2066, 1984.

\bibitem{wapenaar2014}
K.~Wapenaar, J.~Thorbecke, J.~van~der Neut, F.~Broggini, E.~Slob, , and
  R.~Snieder.
\newblock Marchenko imaging.
\newblock {\em Geophysics}, 79:WA39--WA57, 2014.

\bibitem{singh2017}
S.~Singh, R.~Snieder, J.~van~der Neut, Thorbecke J., Slob J., E., and
  K.~Wapenaar.
\newblock Accounting for free-surface multiples in marchenko imaging.
\newblock {\em Geophysics}, 82(1):R19--R30, 2017.

\bibitem{ravasi2016}
M.~Ravasi.
\newblock A method of redatuming geophysical data, 2016.

\bibitem{slob2013}
E.~Slob and K.~Wapenaar.
\newblock Theory for marchenko imaging of marine seismic data with free surface
  multiple elimination.
\newblock {\em 79th EAGE Annual International Conference and Exhibition}, 2017.

\bibitem{ravasi2018}
M.~Ravasi.
\newblock An overview of marchenko-based redatuming: past, present, (and
  future).
\newblock {\em EAGE Annual International Conference and Exhibition}, 2018.

\bibitem{staring2020}
M.~Staring and K.~Wapenaar.
\newblock Three-dimensional marchenko internal multiple attenuation on narrow
  azimuth streamer data of the santos basin, brazil.
\newblock {\em Geophysical Prospecting}, 68:1864--1877, 2020.

\bibitem{bakulin2006}
A.~Bakulin and R.~Calvert.
\newblock {The virtual source method: Theory and case study}.
\newblock {\em Geophysics}, 71(SI139-SI150), 2006.

\bibitem{staring}
M.~Staring, R.~Pereira, H.~Douma, J.~van~der Neut, and K.~Wapenaar.
\newblock Source-receiver marchenko redatuming on field data using an adaptive
  double-focusing method.
\newblock {\em Geophysics}, 83(6):S579–S590, 2018.

\bibitem{cui2018}
T.~Cui, J.~E. Rickett, and I.~Vasconcelos.
\newblock Target-oriented waveform inversion.
\newblock {\em The Journal of the Acoustical Society of America}, 144:1792,
  2018.

\bibitem{gisolf2017}
A.~Gisolf, P.~Haffinger, and P.~Doulgeris.
\newblock Reservoir-oriented wave-equation based seismic avo inversion.
\newblock {\em Interpretation}, 5:1--60, 2017.

\bibitem{minato}
S.~Minato, T.~Matsuoka, T.~Matsuoka, and T.~Tsuji.
\newblock Singular-value decomposition analysis of source illumination in
  seismic interferometry by multidimensional deconvolution.
\newblock {\em Geophysics}, 78(3):Q25--Q34, 2013.

\bibitem{vanderneut2015b}
J.~van~der Neut, J.~Thorbecke, K.~Wapenaar, and E.~Slob.
\newblock Inversion of the multidimensional marchenko equatio.
\newblock {\em 77th EAGE Annual International Conference and Exhibition}, 2015.

\bibitem{wapenaar2020}
K.~Wapenaar and J.~van Ijsseldijk.
\newblock Discrete representations for marchenko imaging of imperfectly sampled
  data.
\newblock {\em Geophysics}, 85(2):A1--A5, 2020.

\end{thebibliography}

\end{document}